# Hybrid EMT-TS Simulation Strategies to Study High Bandwidth MMC-Based HVdc Systems

Yuan Liu[1], Marcelo A. Elizondo[1], Suman Debnath[2], Jingfan Sun[3], Ahmad Tbaileh[1], Yuri V. Makarov[1],
Qiuhua Huang[1], Mallikarjuna R Vallem[1], Harold Kirkham[1], Nader A. Samaan[1]

1, Pacific Northwest National Laboratory
Richland, WA, USA, 99354

2, Oak Ridge National Laboratory
Oak Ridge, TN, USA, 37830

3, Georgia Institute of Technology
Atlanta, GA, USA, 30332

{yuan.liu, marcelo.elizondo, ahmad.tbaileh, yuri.makarov, qiuhua.huang, mallikarjuna.vallem, harold.kirkham, nader.samaan }@pnnl.gov, debnaths@ornl.gov, jingfan@gatech.edu

*Abstract*—Modular multilevel converters (MMCs) are widely used in the design of modern high-voltage direct current (HVdc) transmission system. High-fidelity dynamic models of MMCs-based HVdc system require small simulation time step and can be accurately modeled in electro-magnetic transient (EMT) simulation programs. The EMT program exhibits slow simulation speed and limitation on the size of the model and brings certain challenges to test the high-fidelity HVdc model in system-level simulations. This paper presents the design and implementation of a hybrid simulation framework, which enables the co-simulation of the EMT model of Atlanta-Orlando HVdc line and the transient stability (TS) model of the entire Eastern Interconnection system. This paper also introduces the implementation of two high-fidelity HVdc line models simulated at different time steps and discusses a dedicated method for sizing the buffer areas on both sides of the HVdc line. The simulation results of the two HVdc models with different sizes of buffer areas are presented and compared.

## I. INTRODUCTION

Modern power grid is experiencing a major evolution from *ac* to mixed *ac-dc* transmission systems with decreasing costs and noticeable economic and technical benefits of *dc* technologies. Some of the technical benefits of *dc* technologies include the abilities to interconnect several asynchronous grids, to integrate renewable energy generations, to support underground transmission, to increase the power transfer capability over long distances [1], [2], and to supply loads through variable frequency drives (VFDs) [3] – [5]. The *dc* technologies are also capable of providing ancillary services to enhance the economics and reliability of power systems and optimize the performance of *ac* grids [2].

Several national laboratories and their industry partners launched a jointly research effort to create models and methods to explore and amplify the technical and economic benefits of *dc* technologies in the future grid of United States [1]. This work has developed high-fidelity dynamic models of modular multilevel converters (MMCs) -based high-voltage direct current (HVdc) system [6]. The deliverables also include the development of high-fidelity models and control algorithms of multi-terminal MMC-HVdc systems connecting Eastern Interconnection (EI), Western Electricity Coordinating Council (WECC), and Electric Reliability Council of Texas (ERCOT) [2], [6]. The developed control strategy for the multi-terminal MMC-HVdc system exhibits up to 51.75% improvement in frequency response with the connection of all the three asynchronous power grids in the United States [7]. However, the test of the developed MMC-HVdc system is performed with the aggregated models for EI, ERCOT and WECC grids based on NERC data [8]. These results represent preliminary results that require confirmation tests with the full models. The evaluation on the full model requires the setup of hybrid EMT-TS simulation platform.

Previous research implemented the hybrid simulation platform to represent a significant portion of the system model in a phasor-domain transient stability (TS) simulation program and co-simulate that with a detailed electro-magnetic transient (EMT) model of a small portion containing voltage-source converter (VSC) -HVdc [9],[10]. The communication between the TS and EMT programs is built on a software called E-TRAN Plus [11] that can simultaneously simulate the detailed model using a micro-second timestep in PSCAD [12], and the rest of the system using a milli-second time-step in PSS/E [13]. The PSS/E portion of the system model in [9] utilized an 8-machine 31-bus system, which is a limited-size test system. In addition, the paper by Farsani et al. [9] adopts an empirical approach to size the buffer area, within which a number of *ac* buses on either side of the HVdc line are selected and modeled in PSCAD program.

This paper presents the development of a co-simulation framework to integrate the EMT model of an MMC-HVdc system and the TS model of the external EI system. The characteristics of the EI system are described in detail. Oak Ridge National Laboratory (ORNL) designed an MMC-HVdc model based on hybrid discretization and multi-rate simulation [7]. Two simulation time steps, at 4µs (slow) and 60µs (fast) separately, are implemented for the MMC-HVdc model. This paper also proposes a specific method for sizing the buffer areas on both sides of the HVdc line by selecting the *ac* buses to be modeled in EMT-level simulation program. In this

This paper and the work described were sponsored by the U.S. Department of Energy (DOE) Office of Electricity Delivery and Energy Reliability (OE) Transformers and Advanced Components (TRAC) under the Grid Modernization Laboratory Consortium. The authors would like to thank Dr. Kerry Cheung who leads the DOE TRAC program for providing continued guidance. The authors would also like to thank the contributions of Madhu Chinthavali and Phani Marthi from Oak Ridge National Laboratory.

The Pacific Northwest National Laboratory is operated by Battelle for the U.S. Department of Energy under contract DE-AC05-76RL01830.

paper, PSCAD is used as the EMT-level simulation tool and PSS/E is used as the TS-level simulation program. The software E-TRAN Plus is utilized to provide interfacing between PSCAD and PSS/E.

The following contributions of this paper are highlighted:
- This paper is the first paper that describes the co-simulation of a high-fidelity HVdc transmission line model with realistic EI system while other co-simulation works only consider using limited-size IEEE standard systems [9].
- This paper proposes a VAr injection method to determine the size of buffer areas that are modeled in detailed EMT simulation.

## II. SYSTEM-LEVEL MODEL

An overlay of HVdc macro grid was proposed by Midcontinent Independent System Operator (MISO) with multiple technical and economic benefits [14]. The HVdc macro grid has a maximum transfer capacity of 14.4 GW between the EI and WECC. A version of macro grid is shown in Fig. 1 with power flows from the WECC to the EI.

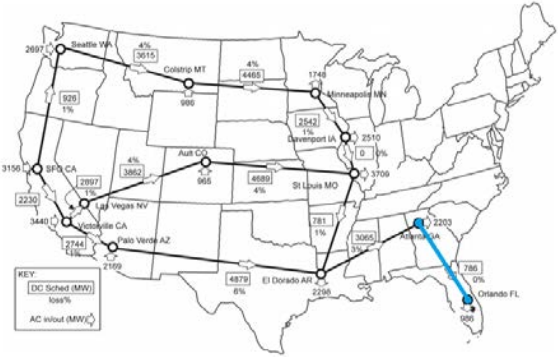

Figure 1. HVdc macro grid design proposed by MISO [1]

The blue HVdc line highlighted in Fig. 1 is the Atlanta-Orlando HVdc line that is modeled in EMT in this study to setup the hybrid simulation platform. This HVdc system is based on MMCs.

Considering the memory limitations in E-Tran Plus and since the HVdc line analyzed in this study is located in EI, WECC part of the model was eliminated and replaced by power injections at the HVdc macro grid terminals. The *ac* interconnection of the EI was modeled by a 2026 Summer-Peak case, provided by the Multi-regional Modeling Working Group of the EI Reliability Assessment Group (ERAG-MMWG). The models used in this study are based on previously developed models in [1]. Key parameters of the final power flow model are shown in TABLE I. The dynamic models of HVdc links except the Atlanta-Orlando line in the HVdc macro grid were represented by the CDC6T HVdc model [13] in PSS/E.

## III. PSCAD MODEL OF MMC-BASED HVDC SYSTEM

### A. MMC Models & Simulation Algorithms

The circuit diagram of a three-phase MMC is shown in Fig. 2. It consists of six arms with *N* series connected submodules (SMs) and an inductor. The basics of operation of the MMC is explained in detail in [15].

TABLE I. SUMMARY OF POWER FLOW NETWORK MODEL PARAMETERS

| Model Characteristics | Quantity |
|---|---|
| Total Generation (GW) | 728 |
| Total Load (GW) | 693 |
| Total Reactive Support (GVAr) | 177 |
| Number of Buses | 78,682 |
| Number of *ac* lines | 99,331 |
| Number of *dc* lines | 68 |
| Number of Generators | 7,829 |
| Number of Loads | 42,730 |

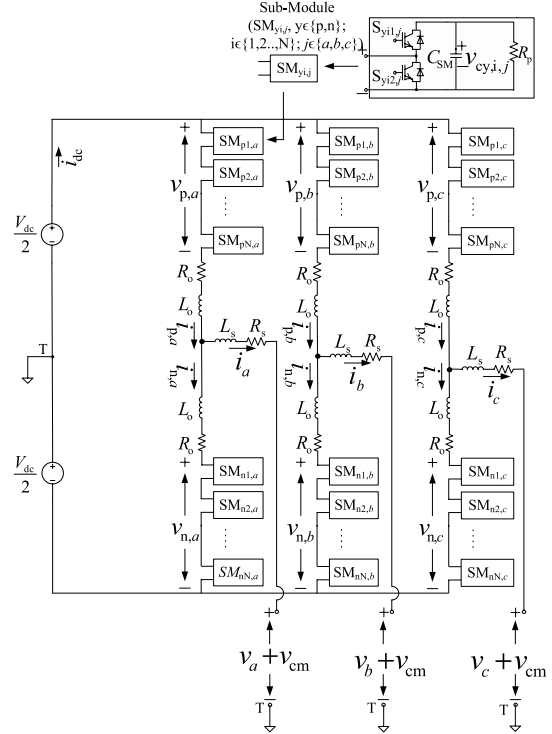

Figure 2: Circuit diagram of MMC

### B. MMC Control Strategies

The hierarchical control of MMC consists of: 1) inner control system to control *ac* grid currents, *dc*-link currents, circulating currents, and SM capacitor voltages, and 2) outer control system to control *ac* voltage and mean of SM capacitor voltages. The inner control system in MMC has been explained in [16] and is summarized in Fig. 3. The *ac* grid currents, *dc*-link currents, and circulating currents' control strategy is shown in Fig. 3. The SM capacitor voltage balancing algorithm is explained in [16] and not repeated here. The outer control system consists of controlling *ac* voltage and mean of SM capacitor voltage, as shown in Fig. 4 [17].

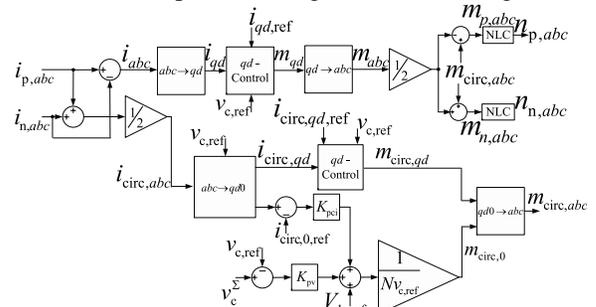

Figure 3: MMC-HVdc arm current control

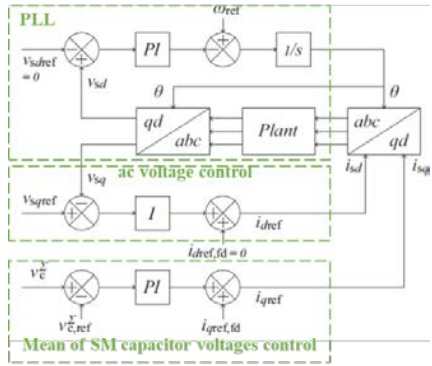

Figure 4: MMC outer controller and PLL

## IV. DESIGN OF BUFFER AREAS

The hybrid simulation in this paper will be performed using E-TRAN Plus [11]. The MMC-HVdc system can perform voltage control functionality. The buffer areas on both rectifier and inverter sides of the HVdc line are identified and modeled in PSCAD to simulate voltage behavior of the model parts close to the HVdc terminals and to increase the accuracy of hybrid simulation. To obtain the buffer areas, a sensitivity-based approach is performed. A reactive power injection is placed around the rectifier or inverter bus and the voltage changes are observed on the surrounding buses. A voltage deviation criterion is defined to determine buses included within the buffer areas. The procedure is illustrated in Fig. 5 and consists of the following steps:

1. Inject a certain amount of reactive power ($\Delta Q$) into an adjacent bus near rectifier or inverter bus.
2. Solve power flow to calculate the voltage variations ($\Delta V$) of adjacent buses in respect to the reactive power injection ($\Delta Q$)
3. Define criteria to rank $\Delta V$ and select buses for which the calculated $\Delta V$ values lie in the specified range.

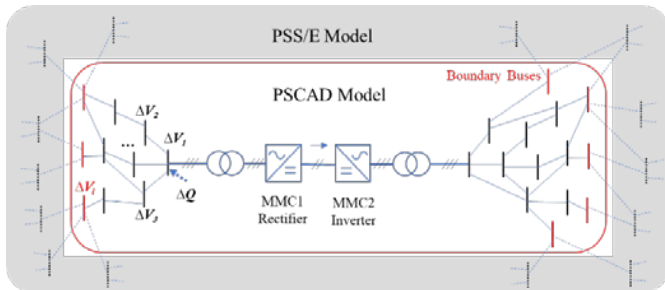

Figure 5. Conceptual illustration of buffer areas selected by reactive power injection and voltage sensitivity method

In this paper, a reactive power injection of 1000 MVAr is used to generate voltage variations. To include a reasonable number of buses, the buses manifesting 1.4% or higher voltage variations are considered within the buffer areas. It is noted that the E-TRAN Plus will add to the buffer areas with several more buses connected to the considered buses through ideal branches. The final number of buses are 50 buses in the rectifier-side buffer area and 12 buses in the inverter-side buffer area. The results are summarized in TABLE II. TABLE II also shows the number of buses for a pair of smaller buffer areas selected based on engineering judgement. The simulation results for the two selections of buffer areas will be compared in Section V.

TABLE II. NUMBER OF BUSES IN THE BUFFER AREAS

| **Large Buffer Areas** *(VAr injection method)* | Total Number of Buses | Number of Boundary Buses |
|---|---|---|
| Rectifier-Side (Atlanta) | 50 | 21 |
| Inverter-Side (Orlando) | 12 | 4 |
| **Small Buffer Areas** *(engineering judgement)* | Total Number of Buses | Number of Boundary Buses |
| Rectifier-Side (Atlanta) | 8 | 4 |
| Inverter-Side (Orlando) | 9 | 4 |

## V. SIMULATION RESULTS

ORNL implemented two PSCAD models of MMC-HVdc line simulated at 4μs (slow model) and 60μs (fast model) separately. In this paper, two selections of buffer areas are proposed in TABLE II. A combination of 4 simulation scenarios is formed to investigate the simulation performances of different MMC-HVdc models in different buffer areas. To compare with the hybrid simulation models, this paper also presents the simulation results of PSCAD equivalent models, for which the MMC-HVdc line and buffer areas are modeled in PSCAD, the external system is modeled as equivalent voltage sources in PSCAD. The simulation scenarios are summarized in TABLE III.

TABLE III. DEPICTION OF SIMULATION SCENARIOS

|  | Fast MMC-HVdc | Slow MMC-HVdc |
|---|---|---|
| Small Buffer Areas | Case 1 | Case 2 |
| Large Buffer Areas | Case 3 | Case 4 |

### A. Contingencies Applied in PSCAD Portion of the System

The simulation lasts for 5 seconds. At t=1.2s, the active power flowing through the HVdc line is ramped up from 0 to 500MW. Two contingencies occurring within the buffer areas are considered when running the PSCAD equivalent model and hybrid simulation model. The first contingency is to trip a 1241 MVAr shunt capacitor close to the rectifier (Atlanta) side of the HVdc line at t=3s and reconnect it at t=3.25s. The second contingency is to apply a 3-phase-to-ground fault at t=4s and clear it at t=4.25s. The voltage, active and reactive power on rectifier and inverter sides are shown in Figs. 6 – 9.

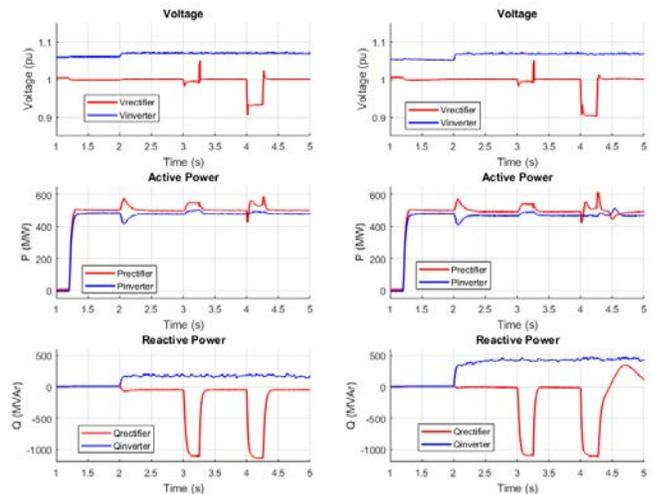

Figure 6. Case 1: **small** buffer area and **fast** MMC-HVdc model (*left*: PSCAD equivalent model, *right*: hybrid simulation model)

Comparing the left and right three subplots in Figs. 6 – 9, PSCAD-PSS/E co-simulation captures power system dynamics more accurately than the PSCAD equivalent model. Following the clearance of the fault at t=4.25s, the system is supposed to experience low-frequency oscillations as demonstrated in many system-level simulation studies [3], [4], [18]. It can be observed from the reactive power subplots in Figs. 6 – 9 that the low-frequency oscillation is captured in the results of hybrid simulation model. Because of the utilization of ideal voltage sources at all boundary buses, the PSCAD equivalent model cannot provide accurate simulation results.

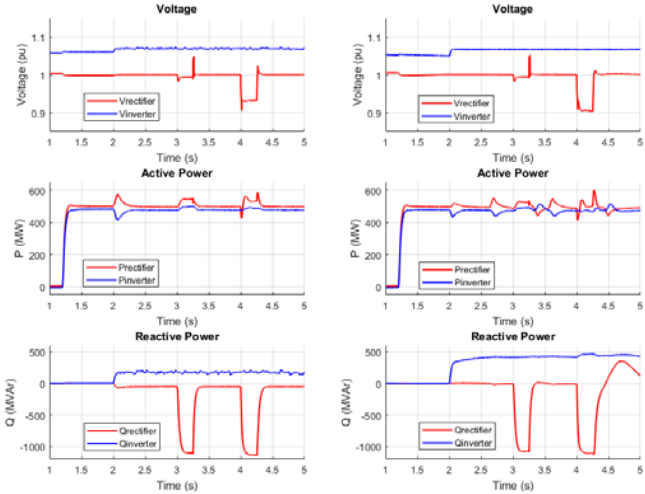

Figure 7. Case 2: **small** buffer area and **slow** MMC-HVdc model (*left*: PSCAD equivalent model, *right*: hybrid simulation model)

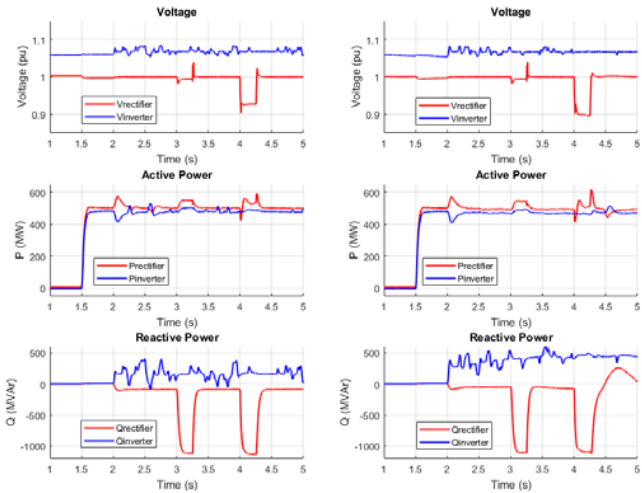

Figure 8. Case 3: **large** buffer area and **fast** MMC-HVdc model (*left*: PSCAD equivalent model, *right*: hybrid simulation model)

Comparing the hybrid simulation results (right three subplots) in Figs. 8 and 9 with that in Figs. 6 and 7, it can be seen that the hybrid simulation model with large buffer areas generates more stable results than that with small buffer areas modeled in PSCAD. For the results of hybrid simulation model with small buffer areas shown in Fig. 7, unwanted swells and sags noticeably appear on the active power curves of the rectifier and inverter. Similar swells and sags are not found in the results of the hybrid simulation models with large buffer areas shown in Figs. 8 and 9. This comparison verifies that the increase in the size of buffer areas improves the stability of the hybrid simulation and accuracy in quantifying the impact of the HVdc control system.

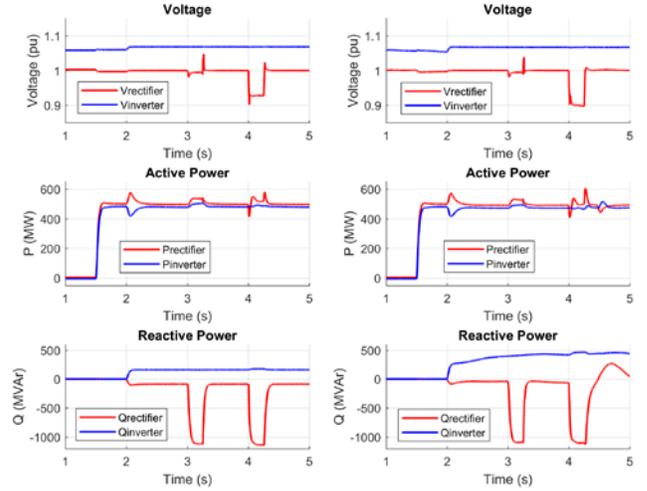

Figure 9. Case 4: **large** buffer area and **slow** MMC-HVdc model (*left*: PSCAD equivalent model, *right*: hybrid simulation model)

Another observation can be drawn from comparing Figs. 7 and 9 with Figs. 6 and 8. The slow MMC-HVdc model, simulated at 4μs, renders smoother (less noisy) results than the fast MMC-HVdc model that is simulated at 60μs. The reason for this observation can be attributed to the slower sampling assumed in the slow MMC-HVdc model that interferes with the control system's response. This phenomenon is more pronounced in the case with the larger buffer zone model (see Fig. 9), indicating the loss of fidelity in the models with a smaller buffer zone (see Fig. 7).

The simulation performances for the four cases are presented in TABLE IV. The simulation cases tabulated in TABLE III are completed in a laptop with a 64-bit operating system, Intel Core i7-6820HQ CPU at 2.7 0GHz and 16GB RAM. It can be concluded from TABLE IV that ORNL's fast HVdc line model results in up to 3.9 times faster computation time than slow model.

TABLE IV. COMPUTATION PERFORMANCE FOR 5-SECOND SIMULATION LENGTH

| Buffer area size | Simulation model | PSCAD time step - ORNL's line model | PSS/E time step | Simulation time | Speedup |
|---|---|---|---|---|---|
| Small | PSCAD with equivalent | 60 μs – Fast | - | 247 s | 2.4x |
| | | 4 μs – Slow | - | 610 s | |
| | PSCAD-PSS/E co-simulation | 60 μs – Fast | 4.16 ms | 287 s | 2.4x |
| | | 4 μs – Slow | 4.16 ms | 706 s | |
| Large | PSCAD with equivalent | 60 μs – Fast | - | 303 s | 3.9x |
| | | 4 μs – Slow | - | 1189 | |
| | PSCAD-PSS/E co-simulation | 60 us – Fast | 4.16 ms | 391 s | 3.4x |
| | | 4 us – Slow | 4.16 ms | 1351 | |

B. *Contingencies Applied in PSS/E Portion of the System*

This scenario considers a contingency applied on the PSS/E side of the hybrid simulation model. Four adjacent generators with a total generation of 3.512 GW are tripped offline at t=3s on the PSS/E side. The rectifier- and inverter-side voltages, active and reactive powers are monitored on the PSCAD side for the 4 scenarios described in TABLE III. The results are shown in Fig. 10.

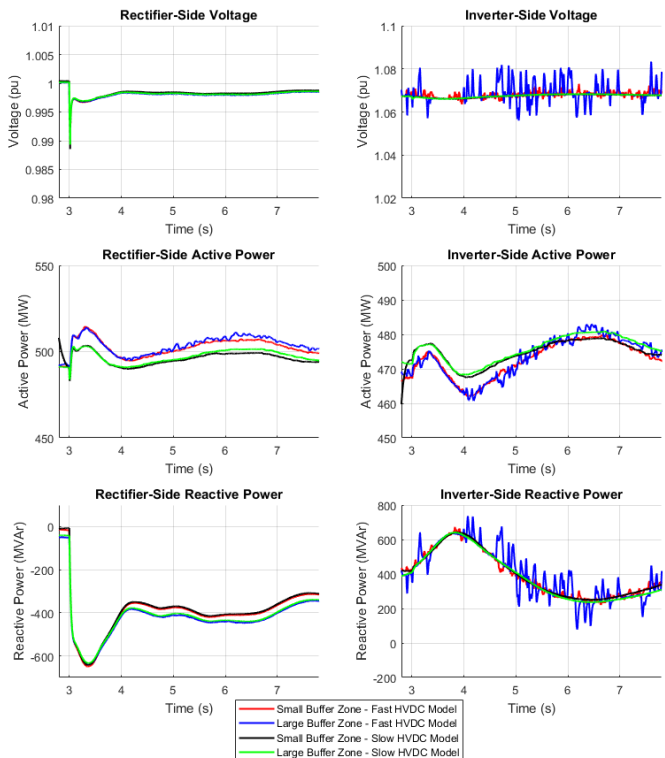

Figure 10. HVdc rectifier- and inverter- side voltages, active and reactive powers monitored in PSCAD in response to generator tripping on PSS/E side

It can be seen from Fig. 10 that all the four models generate similar voltage and power trajectories. However, high-frequency noise is obviously seen on the voltage and power curves of the fast MMC-HVdc models with small and large buffer areas. Based on the observations from Figs. 6 – 9, it can be concluded that the slow MMC-HVdc model with large buffer areas is able to generate the most credible and stable simulation results. However, in preliminary studies where speed is important, one may begin with a small buffer zone and fast MMC-HVdc model, which also provides relatively reliable results (see Fig. 6 and Fig. 10).

## VI. CONCLUSIONS

This paper presents the development of the hybrid EMT-TS simulation framework to study the responses of high-fidelity HVdc line in the full North American EI system. The PSCAD (or EMT) model of the MMC-based HVdc system is implemented based on fast and slow approaches and its control system is briefly described in this paper. The sizes of the interfacing areas between the HVdc line and EI system, known as buffer area, are determined by VAr injection techniques. It can be concluded from the studies in this paper that the size of buffer area in the hybrid simulation of HVdc line impacts the accuracy and stability of the simulation performances. A large buffer area is essential for accurate EMT-TS co-simulation with a high-fidelity HVdc system. Though the simulation speed is compromised, a slow MMC-HVdc model is a preferable choice to provide reliable simulation results. In preliminary studies where speed is important, one may begin with a small buffer zone and fast MMC-HVdc model.